\documentclass[conference]{IEEEtran}

\usepackage{graphicx}  

\usepackage{psfrag}    

\usepackage{subfigure} 

\usepackage{amsmath}   

\usepackage{amssymb}
\usepackage{amsfonts}
\usepackage{latexsym}
\usepackage{mathrsfs}

\begin{document}

\title{An OFDM-CDMA scheme for High Data Rate UWB applications}

\author{\authorblockN{Emeric Gu\'eguen, Matthieu Crussi\`ere, Jean-Fran\c cois H\'elard}
\authorblockA{Electronics and Telecommunications Institute of Rennes\\
IETR/INSA, 20 Av. des Buttes de Co\"esmes\\
35043 Rennes Cedex, France\\
\{emeric.gueguen, matthieu.crussiere, jean-francois.helard\}@insa-rennes.fr}
}


%


\maketitle

\begin{abstract}

In this paper, we investigate a new waveform for UWB systems obtained by the combination of Orthogonal Frequency Division Multiplex (OFDM) and Code Division Multiple Access (CDMA). The proposed system, called Spread Spectrum - Multi-Carrier - Multiple Access (SS-MC-MA) turns out to be a judicious solution to combat frequency selectivity and narrowband interferers, and to manage the coexistence of several users and piconets. It is shown that the addition of a degree of freedom brought by the spreading component of SS-MC-MA allows to optimize jointly the assignment of the number of used codes and coding rates in order to make the system more robust. Through simulations, it is demonstrated that the new system can outperform Multi-Band OFDM Alliance (MBOA) for low data rates and is able to provide wider range of rates. \\


\textit{Index Terms}--- UWB, MB-OFDM, MC-CDMA, SS-MC-MA.


\end{abstract}


%
\IEEEpeerreviewmaketitle

\section{Introduction}

Ultra Wide Band (UWB) radio systems are today acknowledged as high potential solutions for Wireless Personal Area Networks (WPAN). The novelty of these systems lies in the possibility of non regulated access to the spectral resource leading to a flexible use of the radio channel for an important number of applications. 

In 2002, the Federal Communications Commission (FCC) regulated UWB systems by imposing a spectral mask to limit the transmission power~\cite{FCC_report}. According to the FCC regulation, a signal must have a minimum bandwidth of 500 MHz or a bandwidth to central frequency ratio above 0.2 to be considered as UWB. The power spectral density (PSD) should also not exceed -41.3 dBm/MHz. The UWB channel, running from 3.1 to 10.6 GHz, is frequency selective and considered as almost invariant in time. Proposed UWB systems must not disturb existing narrowband systems, such as Wireless Local Area Network (WLAN) 802.11a standard at 5 GHz for example.

Following that, the Institute of Electrical and Electronics Engineers (IEEE) initiated a standardization process, referred to as task group 802.15.3a, to define a high data rate physical layer for WPAN. The last three years have hereby seen the emergence and the confrontation of two approaches: a \emph{pulse radio} solution using Direct Sequence - Code Division Multiple Access (DS-CDMA) ternary codes supported by UWB Forum, and a \emph{multi-carrier multi-band} solution based on Orthogonal Frequency Division Multiplex (OFDM). The latter has been proposed by the Multi-Band OFDM Alliance (MBOA) consortium and is currently promoted by the main actors of the general public and component industries~\cite{Batra_TG3a}.

In section II, this paper presents the main parameters of the MBOA solution. After a critical analysis of this solution section III the interest of adding a CDMA component to the MB-OFDM waveform. In section IV, the principle of the new waveform called Spread Spectrum - Multi-Carrier - Multiple Access (SS-MC-MA) is presented. Particularly, SS-MC-MA offers for the future WPAN good performance and great flexibility for the resource allocation between users of a same piconet. This new UWB system is then described in section V and the comparison of the performances obtained with the MBOA solution and the SS-MC-MA system is presented in section VI. Finally, section VII concludes the paper.

\section{The MBOA solution}

\begin{figure}
	\centering
	\includegraphics[width=3.5in]{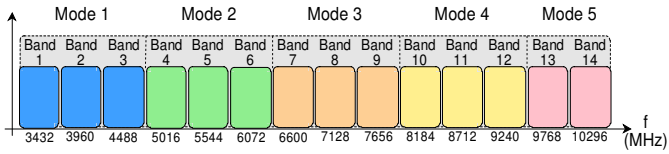}
	\caption{Channels distribution for MBOA solution}
	\label{MBOA_channel_cutting}
\end{figure}

The MBOA consortium proposes to divide the available band into 14 sub-bands of 528 MHz, as illustrated in Fig.~\ref{MBOA_channel_cutting}. Each of these sub-bands allows the transmission of an OFDM signal, obtained from a 128-point Inverse Fast Fourier Transform (IFFT). Most of the UWB studies focus on the first mode which clusters the first three sub-bands from 3.1 to 4.8 GHz. The multi-user management within a piconet is based on Time Division Multiple Access (TDMA) by using a Time-Frequency Code (TFC). At a given moment, each user then occupies one of the three sub-bands of mode 1~\cite{Batra_TG3a}. The signal, sampled during the analog-to-digital conversion, has a limited bandwidth of 500 MHz, leading to low-cost components and power consumption reduction. However, the use of TFC translates into frequency hopping from one sub-band to another at each OFDM symbol. Hence, each user benefits from the frequency diversity brought by the three sub-bands in mode 1. In addition, considering that each user occupies a given sub-band only one third of the time, it is possible to optimize the transmitted power while respecting the PSD mask imposed by the FCC. Lastly, it is also advised to plan the cohabitation of 4 piconets in a same environment. 

Transmitted data rates in each sub-band essentially depend on the coding rate, as the modulation applied to the different subcarriers of the OFDM multiplex is a quadrature phase-shift keying (QPSK). Data rates, from 53.3 to 480 Mbit/s, are listed in Table~\ref{MBOA_data_rates}. 
For certain modes, each complex symbol and its conjugate symmetric are transmitted into the same OFDM symbol by one subcarrier and its "mirror" subcarrier respectively. This way, the frequency diversity is exploited into each sub-band at the cost of a division by 2 of the useful transmitted data rate. Moreover, for modes corresponding to data rates from 53.3 Mbit/s to 200 Mbit/s, a time spreading of 2 is applied. It consists in the transmission of the same information during 2 consecutive OFDM symbols in order to take advantage of a better frequency diversity, due to the joint application of the TFC. All OFDM parameters of the MBOA solution are detailed in~\cite{Batra_TG3a}. 

\begin{table}
	\renewcommand{\arraystretch}{1.3}
	\caption{MBOA data rates}
	\label{MBOA_data_rates}
	\centering
	\begin{tabular}{c c c c c c}
	\hline
	\bfseries Data &  \bfseries Modulation &  \bfseries Coding &  \bfseries Conjugate &  \bfseries Time &  \bfseries Coded \\
	\bfseries rate  &    & \bfseries rate &  \bfseries symmetric  & \bfseries spreading  & \bfseries bits per  \\
	\bfseries 	(Mbit/s) &		 & 	\bfseries (\textit{R})  & \bfseries input &  \bfseries factor	 & \bfseries OFDM \\
	  &  		 & 			& \bfseries to IFFT  & \bfseries (TSF)  & \bfseries symbol \\
	\hline
	53.3 & QPSK & 1/3 	& Yes & 2	& 100 \\
	80   & QPSK &	1/2 	& Yes & 2	& 100 \\
	110  & QPSK	&	11/32 & No 	& 2	& 200 \\
	160  & QPSK	&	1/2 	& No 	& 2	& 200 \\
	200  & QPSK	&	5/8 	& No	& 2	& 200 \\
	320  & QPSK	&	1/2 	& No 	& 1	& 200 \\
	400  & QPSK	&	5/8 	& No 	& 1	& 200 \\
	480  & QPSK	&	3/4 	& No 	& 1 & 200 \\
	\hline
	\end{tabular}
\end{table}

One of the main difference compared to a classical OFDM system is the use of a zero padding (ZP) guard interval instead of the traditional cyclic prefix (CP). Indeed, CP is replaced by trailing zeros. Details of this operation are well explained in~\cite{Muquet_ZP}. ZP allows one to obtain a spectrum with fewer ripples in the useful band than with a traditional CP. Thus the signal can take the exact shape of the PSD mask~\cite{Batra_TG3a}. 

To summarize, the MBOA solution offers some advantages for high data rate UWB applications, such as the signal robustness against the channel selectivity and the efficient exploitation of the signal energy received within the prefix duration. The main argument of multi-carrier modulation in general is often quoted in favour of the MBOA solution, when one compares it with the competitive DS-CDMA solution. The latter can actually hardly make use of all the received energy, the number of the RAKE fingers being compulsorily limited for complexity reasons. However, the MBOA solution is relatively limited in a multi-user and multi-piconet context. Particularly, when the only three first sub-bands of the first mode are considered, conflicts immediately appear at the addition of a fourth user within a piconet, whereas scenarios going up to 6 simultaneous users have classically to be considered.

\section{The CDMA component add interest}

\begin{figure}
	\centering
	\includegraphics[width=3.2in]{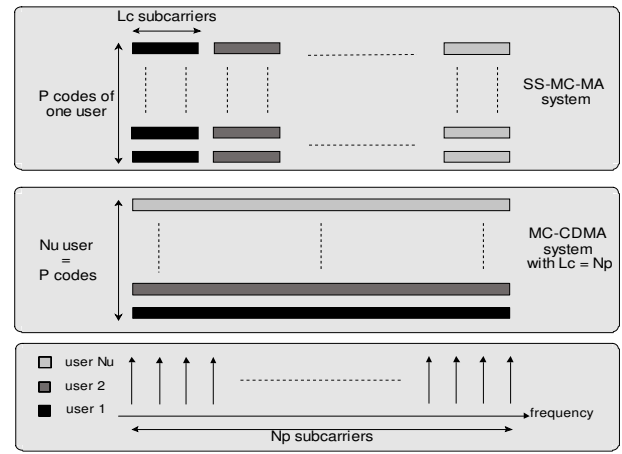}
	\caption{Data distribution of different users for MC-CDMA and SS-MC-MA systems}
	\label{MC-CDMA_and_SS-MC-MA}
\end{figure}

Recent studies have proposed to add a CDMA component to the MBOA solution in order to improve the system robustness or the resource sharing between several users~\cite{Y-B_Park}. This spreading component essentially allows to organize the access of several users to a common resource. Taking into account the UWB channel characteristics, frequency selectivity and slow time variations in indoor environment, spreading is generally performed along the frequency axis, leading to a Multi Carrier-CDMA (MC-CDMA) waveform. The symbols of all users are transmitted by all the subcarriers as depicted in Fig.~\ref{MC-CDMA_and_SS-MC-MA}, the spreading code length $L_c$ being lower or equal to the subcarrier number $N_p$ of the OFDM multiplex. 


Compared to the "traditional" MBOA solution, and beyond a greater facility in the resource sharing, the MC-CDMA system also presents a better robustness against channel frequency selectivity (\cite{Y-B_Park},~\cite{M_Schmidt}) and improves the UWB signal robustness against narrowband interferences. This last point is fundamental for uncontrolled access to the spectral resource. In~\cite{Y-B_Park} however, authors suggest to use an MC-CDMA signal with a bandwidth $B_w$ = 1.58 GHz, equivalent to 3 sub-bands of the MBOA signal, which leads to an highly increase of the sampling frequency of the analog-to-digital conversion.


\section{A new waveform for multi-band UWB : the SS-MC-MA}

By applying a spreading code and a multi-access component, we propose in this paper an SS-MC-MA waveform~\cite{L_Cariou}, which is new for UWB applications and offers better performance and more flexibility in the resource management. 

\subsection{SS-MC-MA principle}

SS-MC-MA can be viewed as a multi-block system compared to the classical MC-CDMA system, since the spectrum is divided into "blocks" of several subcarriers. The SS-MC-MA solution, illustrated in Fig.~\ref{MC-CDMA_and_SS-MC-MA}, consists in assigning to each user a specific set of subcarriers according to a Frequency Division Multiple Access (FDMA) approach. Code dimension $L_c$ can then be exploited for an adaptive resource optimization and sharing (modulation type, data rate, \ldots). Spreading in the frequency domain leads to diversity gain and, as it is the case of MC-CDMA, improves the signal robustness against narrowband interferers. With an SS-MC-MA signal, symbols are transmitted simultaneously on a specific subset of subcarriers by the same user and undergo the same distortions. Self-interference (SI) which then replaces the Multiple Access Interference (MAI) obtained with MC-CDMA signals, can be easily compensated for by mono-user detection with only one complex coefficient per subcarrier.

%

\subsection{SS-MC-MA advantages}

Let us consider the case of the MBOA standard in mode 1 (Fig.~\ref{MBOA_channel_cutting}).

\subsubsection{Case of three or less users}
The SS-MC-MA system allows the allocation of a 528 MHz sub-band for each user. This system offers the same performance and advantages as MC-CDMA to which a simplicity is added for the channel estimation in reception. In fact, with SS-MC-MA a given subcarrier is distorted by only one channel, the one of the user associated with this subcarrier. At contrary, with an MC-CDMA system, each subcarrier is corrupted by the different channels of different users, which increases considerably channel estimation complexity. In that case, each user has to estimate the response of many channels all over the total available bandwidth.
\subsubsection{Case of more than three users}
In the MBOA solution, conflicts appear from 4 users and could cause information losses. In the SS-MC-MA case, the code dimension could be exploited to share a same 528 MHz sub-band between 2 or even 3 users if necessary. In that case, the generated signal within a given block corresponds to an MC-CDMA signal, but with a limited number of users per block (2 or even 3).

More generally, in a multi-piconet context, the possibility of easily modifying the number of spreading codes assigned to a given user in a given piconet, allows the SS-MC-MA scheme to offer a more flexible and efficient dynamic resource sharing than the MBOA solution does.

\section{The new UWB system}

\subsection{System studied}

\begin{figure}
	\centering
	\includegraphics[width=3.5in]{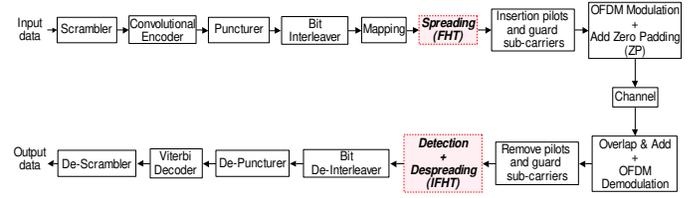}
	\caption{MBOA transmission chain (SS-MC-MA in broken lines)}
	\label{MBOA_SSMCMA_chain}
\end{figure}

The proposed system is based on the MBOA solution. Fig.~\ref{MBOA_SSMCMA_chain} introduces the MBOA transmission chain in continuous lines and, in dashed lines, the functions that are added to obtain an SS-MC-MA waveform. These functions are mainly the Hadamard Transform ("Fast Hadamard Transform": FHT) at the transmitter and the inverse transform (IFHT) at the receiver. In addition, Minimum Mean Square Error (MMSE) single user detection is applied.

The spreading factor $L_c$ is chosen equal to $16$ and the number of useful subcarriers is reduced from $100$ to $6\times16 = 96$ for each OFDM symbol. This means that 4 more guard subcarriers are added.

\subsection{Signals expression}

In the MBOA solution case, the signal generated at the output of the IFFT expresses:

\begin{equation}
	\label{Signal_MBOA}
		\begin{array}{ll}
			S_\textsc{ofdm}(t) = \sum\limits_{i=-\infty}^{+\infty} \ \sum\limits_{n=-N_\textsc{st} /2}^{n=+N_\textsc{st} /2} & X_n(i)p_c(t-iT_\textsc{cp})\\
			& \times \ e^{j2\pi n\Delta_\textsc{f}(t-iT_\textsc{cp})}
		\end{array}
\end{equation}

\noindent where $\Delta_\textsc{f}$, $N_\textsc{st}$ and $T_\textsc{cp}$ represent the subcarriers spacing, the total number of used subcarriers and the spacing between two consecutive OFDM symbols, respectively. $X_n(i)$ is a complex symbol belonging to a QPSK constellation and is transmitted by subcarrier $n$ during the $i$th OFDM symbol. It represents a data, a pilot or a reference symbol. $p_c(t)$ is a rectangular window defined by:

\begin{equation}
	p_c(t) = \left\{ \begin{array}{ll}
	1 & \textrm{$0\leq t\leq T_\textsc{fft}$}\\
	0 & \textrm{$T_\textsc{fft}\leq t\leq T_\textsc{fft}+T_\textsc{cp}+T_\textsc{gi}$}
		\end{array} \right.
	\label{rect_window}
\end{equation}

In the SS-MC-MA case, complex symbols are converted into $P$ parallel symbols $D_l(i)$ (with $P \leq L_c$) which are transmitted by the same $L_c$ subcarriers. $P$ then represents the load, and is namely equal to $L_c$ in the full load case and to $L_c/2$ in the half load case. The waveform is the same as previously, but the complex symbol $X_m(i)$ which is transmitted by the $m$th subcarrier ($m$ varying from $1$ to $L_c=16$, with $m=n$ modulo($16$)) of a block of $L_c=16$ subcarriers bound by the same spreading codes of length $L_c$ can be express by:

\begin{equation}
	X_m(i) = \sum_{l=1}^{P} D_l(i)c_{l,m}
	\label{sym_cplx}
\end{equation}

\noindent where $C_l = [c_{l,1} ... \ c_{l,m} ... \ c_{l,L}]$ is the $l$th Walsh-Hadamard orthogonal spreading code and $D_l(i)$ represents the $P$ complex symbols, belonging to a QPSK constellation and which are transmitted by the block of $L_c$ subcarriers considered during OFDM symbol $i$. In reception, as in the classical OFDM system case, mono-user detection is simply realized at the output of the FFT by one complex multiplication per subcarrier. MMSE detection technique is considered leading to coefficients:

\begin{equation}
g_{n,i} = \frac{h_{n,i}^{*}}{\left| h_{n,i}\right|^2+\frac{1}{\gamma_{n,i}}}
\label{MMSE}
\end{equation}

\noindent where $h_{n,i}$ and $\gamma_{n,i}$ represent the complex channel response and the signal to noise ratio for subcarrier $n$ of symbol $i$ respectively.

\subsection{UWB channel modeling}

The channel model used in the MBOA and SS-MC-MA chain is the one adopted by the IEEE 802.15.3a channel modelling sub-committee to evaluate the UWB physical layer performance. This channel model results from Saleh-Valenzuela model for indoor application~\cite{Saleh_Valenzuela}. This ray based model takes into account clusters phenomena highlighted during channel measurements. Mathematically, the impulse response of the multipath model is given by:

\begin{equation}
	\label{channel_impulse_response}
	h_k(t) = X_k \sum\limits_{m=0}^{M_k} \sum\limits_{p=0}^{P_k} \alpha_k(m,p)\delta\left(t-T_k(m)-\tau_k(m,p)\right)
\end{equation}

\noindent where $X_k$ is the log-normal shadowing for the $k$th channel realization, $\alpha_k(m,p)$ and $\tau_k(m,p)$ are the gain and the delay of path $p$ of cluster $m$ respectively, and $T_k(m)$ is the delay of cluster $m$. Table \ref{Channel_CMi} gives the mean excess delay $\tau_m$ and the root mean square delay spread $\tau_{rms}$ for the 4 channel models CM$i$. In the Line Of Sight (LOS) configuration, transmitter and receiver antennas are in direct visibility, contrary to the Non Line Of Sight (NLOS) configuration. 

\begin{table}
	\renewcommand{\arraystretch}{1.3}
	\caption{Characteristics of wideband channels CM\textit{i}}
	\label{Channel_CMi}
	\centering
	\begin{tabular}{c c c c c}
	\hline
	  & \bfseries CM1 & \bfseries CM2 & \bfseries CM3 & \bfseries CM4\\
	\hline
	\bfseries Mean excess delay (ns): $\tau_m$  & 5.05 		& 10.38 	& 14.18 	&  			\\
	\bfseries RMS delay spread : $\tau_{rms}$ 	& 5.28 		& 8.03 		& 14.28  	& 25		\\
	\bfseries Distance (m) 											& $<$ 4 	& $<$ 4 	& 4 - 10  & 10		\\
	\bfseries LOS/NLOS 													& LOS 		& NLOS 		& NLOS 		& NLOS	\\
	\hline
	\end{tabular}
\end{table}

The channel is modelled in the time domain, and is normalized in mean energy for each realization. 100 different realizations are used for each CM$i$, one realization being applied along a whole frame duration. 


\subsection{Choice of the spreading code}

To reduce the SI, spreading sequences or codes are chosen orthogonal. In presence of a multipath channel, the orthogonality between signals is broken and an SI term appears. It is shown that some combinations of spreading codes could increase or decrease the SI power. Thus, \cite{Snobilet_special_issue} proposed a method to minimize this SI which consists of a judicious subgroup selection of $N_u$ spreading sequences. These methods are used in this study to select the spreading codes. 

\section{Systems performances}

\begin{figure}
	\centering
	\includegraphics[width=0.45\textwidth,height=0.6\linewidth,angle=0]{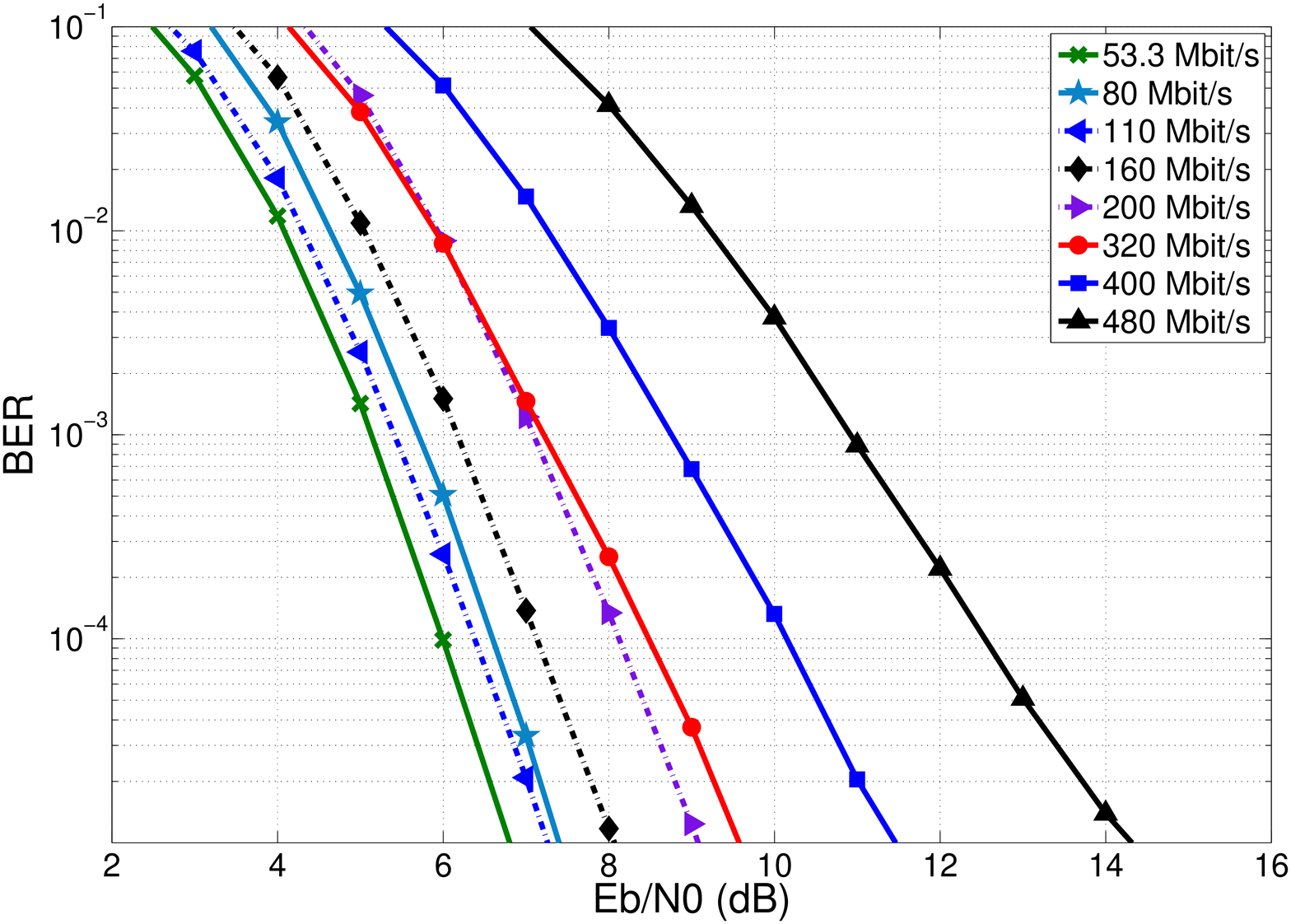}
	\caption{Performance of the MBOA system with channel CM1}
	\label{fig_MBOA_CM1}
\end{figure}

Firstly, the performance of the MBOA system has been estimated for UWB channel CM1. In simulations, frames of $150$ OFDM symbols are used, and one different channel realization is applied on each new frame. Fig.~\ref{fig_MBOA_CM1} exhibits the results obtained in the ideal case of perfect channel estimation for rates ranged from 53.3 to 480 Mbit/s. It appears that the rate couples 53.3-110 Mbit/s and 80-160 Mbit/s give closed results. This shows that the conjugate symmetric does not improve performance as well as expected for data rates 53.3 and 80 Mbit/s. 

Similar results not presented herein have been obtained with channel CM4 leading to the same conclusions, except that a performance floor can appear at high SNR for high data rates, i.e. for high channel code rates. This performance degradation can simply be understood noticing that the GI length is smaller than the delay spread of CM4 \cite{Maret_Siaud_ECPS}.

Let us now focus on the performance obtained with the proposed SS-MC-MA system on channel CM1. The simulation configurations are similar to those used with MBOA, even if the total number of subcarriers $N_{ST}$ is henceforth 118 instead of 122 to take into account the spreading component, the guard and pilot subcarriers. Consequently, the transmission bandwidth becomes equal to 490.87 MHz instead of 507.37 MHz. Via the assignment of a given number $P$ of spreading codes and the choice of the coding rate, it is possible to obtain many different data rates. Table~\ref{SS-MC-MA_data_rate} introduces the code number/coding rate pairs that lead to SS-MC-MA data rates very close to the MBOA ones. As it will be detailed in the following, these pairs correspond to the best choices, among other possible solutions, in term of BER results. The additional degree of freedom brought by the ability to select a given number of spreading codes allows to reach the target rates without applying the conjugate symmetric function nor the time spreading.

\begin{table}
	\renewcommand{\arraystretch}{1.3}
	\caption{Possible Data Rates with SS-MC-MA}
	\label{SS-MC-MA_data_rate}
	\centering
	\begin{tabular}{c c c c c}
	\hline
	\bfseries Data rate & \bfseries Modulation & \bfseries Coding Rate & \bfseries Load & \bfseries Coded bit \\
	\bfseries (Mbit/s) & & \bfseries (\textit{R}) & \bfseries \textit{P} & \bfseries per symbol\\
	\hline
	51.2  & QPSK & 1/3 & 4  & 48 	\\
	76.7  & QPSK & 1/3 & 6  & 72 	\\
	115.1 & QPSK & 1/3 & 9  & 108 \\
	153.6 & QPSK & 1/3 & 12 & 144 \\
	192 	& QPSK & 1/2 & 10	& 120 \\
	307 	& QPSK & 1/2 & 16 & 192	\\
	409 	& QPSK & 2/3 & 16 & 192	\\
	460 	& QPSK & 3/4 & 16 & 192	\\
	\hline
	\end{tabular}
\end{table}

\begin{figure}
	\centering
	\includegraphics[width=0.45\textwidth,height=0.6\linewidth,angle=0]{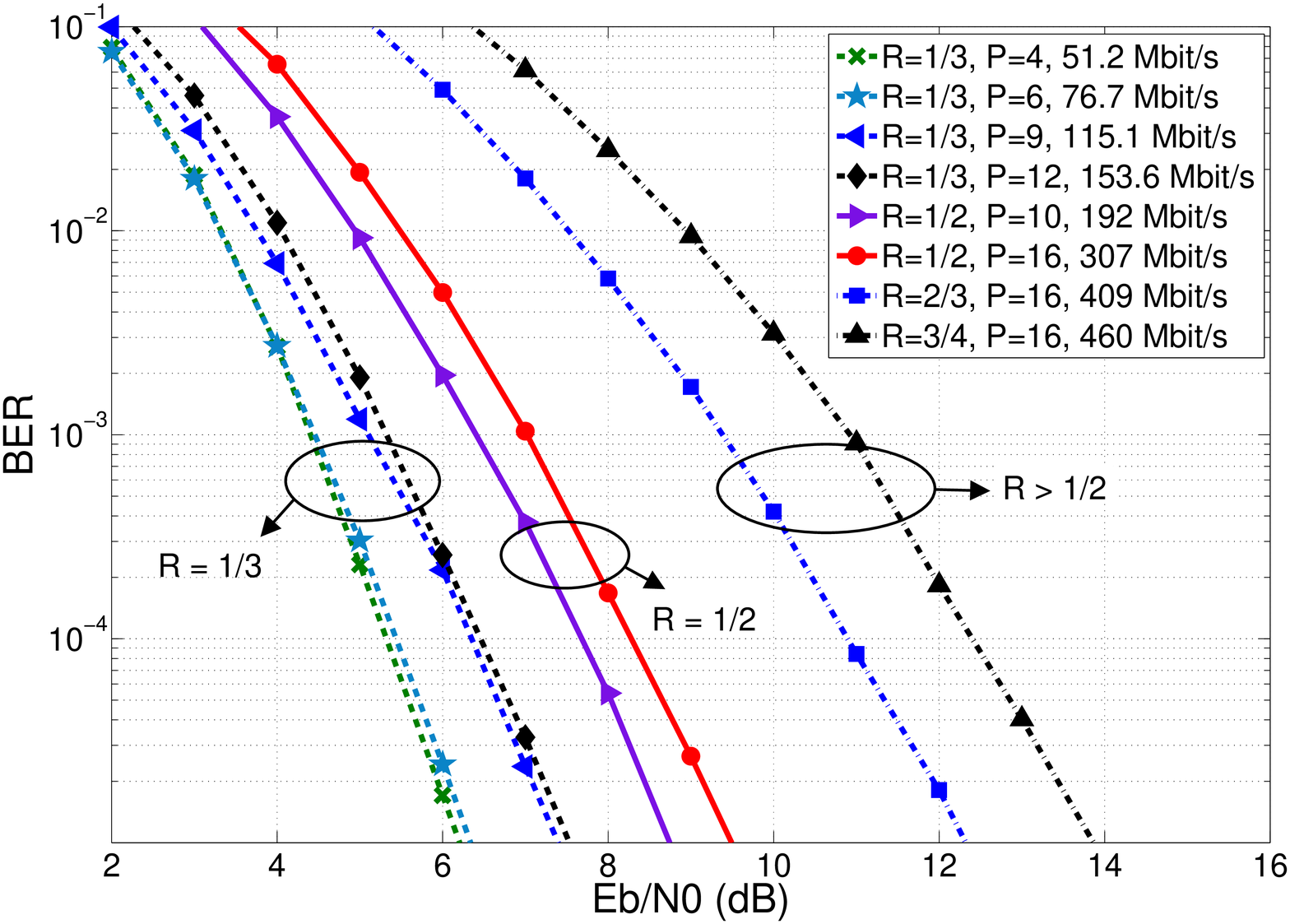}
	\caption{Performance of the SS-MC-MA system with channel CM1}
	\label{fig_SS-MC-MA_CM1}
\end{figure}

Fig.~\ref{fig_SS-MC-MA_CM1} exhibits the results obtained for the data rates of table~\ref{SS-MC-MA_data_rate}. It is shown that SS-MC-MA with R=$1/3$ clearly outperforms MBOA, whereas the two systems MBOA and SS-MC-MA with R=$1/2$ gives quasi similar results. This behavior is consistent with the conclusions already drawn in other studies \cite{Ibars_Globecom01} that tends to show that hybrid OFDM/CDMA systems perform at best for low coding rates (e.g. $1/3$). In that case, the diversity brought by the spreading function leads to a substantial performance increase while the SI effect remains limited. At the contrary, for high coding rates, the diversity gain is almost completely reduced by the performance degradation caused by the SI. 

\begin{figure}
	\centering
	\includegraphics[width=0.45\textwidth,height=0.53\linewidth,angle=0]{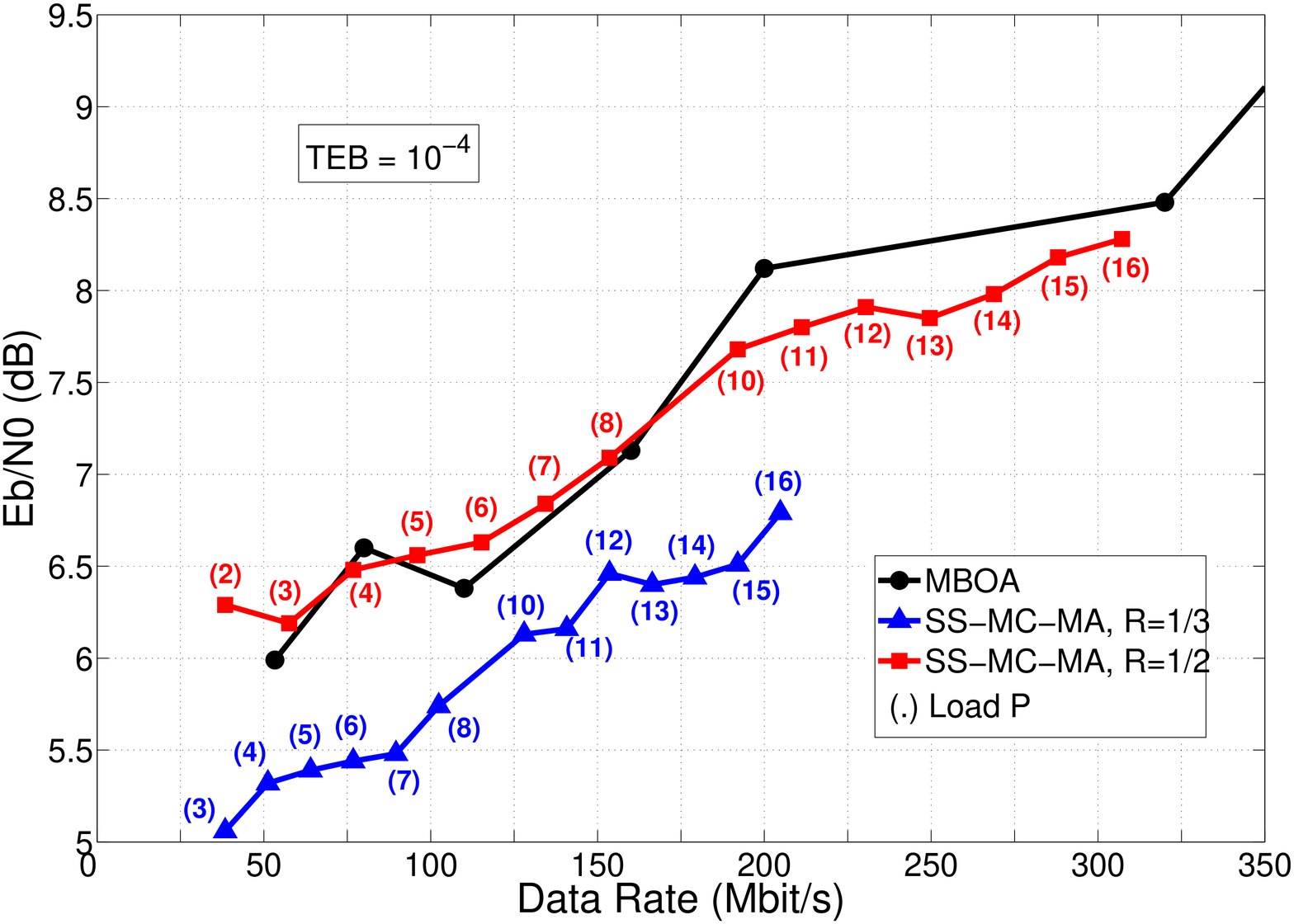}
	\caption{Required $E_b/N_0$ to obtain a BER=$10^{-4}$ on CM1}
	\label{fig_compare_MBOA_SS-MC-MA_CM1}
\end{figure}

To emphasize on the performance gain at low data rate, Fig.~\ref{fig_compare_MBOA_SS-MC-MA_CM1} exhibits comparative results of the two systems. The plotted curves give for each targeted data rates the $E_b/N_0$ required to obtain a BER equal to $10^{-4}$. Two coding rates, $1/2$ and $1/3$ are considered with SS-MC-MA, and the number $P$ of spreading codes is mentioned near each marker. Firstly, it is clear that a coding rate of $1/3$ should be exploited with SS-MC-MA for data rates lower than 200 Mbit/s as already introduced in table~\ref{SS-MC-MA_data_rate}. As expected from the previous conclusions, SS-MC-MA (R=$1/3$) outperforms MBOA for data rates lower than 200 Mbit/s. These results essentially highlight that the MBOA solution based on TSF and conjugate symmetric is not efficient. On the other hand, for R=$1/2$ the proposed system performs very close to MBOA, however providing a wider range of rates due to the high flexibility brought by the joint assignment of the number of used codes and coding rates.

\section{Conclusion}

In this paper, we proposed a new waveform for UWB systems based on the combination of OFDM and CDMA which is called SS-MC-MA. The main interest of this solution is a better resource allocation in multi-user and multi-piconet as well as a better robustness against frequency selectivity and narrowband interferers. With SS-MC-MA, the addition of a degree of freedom brought by the spreading component allows to optimize jointly the assignment of the number of used codes and coding rates. Particularly, it has been shown that the optimized new system outperforms MBOA for low data rates and is able to provide wider range of rates. These improvements could be obtained without increasing the system complexity in comparison with the reference MBOA solution since the transmission chain only requires the addition of an Hadamard transform function. 


\section*{Acknowledgment}
The authors would like to thank France T\'el\'ecom R$\&$D/RESA/BWA which supports this study within the contract 461 365 82.



%

\end{document}